# Computational Skills by Stealth in Secondary School Data Science

Wesley Burr[a], Fanny Chevalier[b], Christopher Collins[c], Alison L Gibbs[d]*, Raymond Ng[e], Chris Wild[f]

[a]Department of Mathematics, Trent University, [b]Departments of Computer Science and Statistical Sciences, University of Toronto; [c]Faculty of Science, Ontario Tech University; [d]Department of Statistical Sciences, University of Toronto: [e]Department of Computer Science, University of British Columbia; [f]Department of Statistics, University of Auckland

*Contact: Alison L Gibbs, alison.gibbs@utoronto.ca, Department of Statistical Sciences, University of Toronto, Toronto, Ontario, Canada M5S 1A1

# Computational Skills by Stealth in Secondary School Data Science

ABSTRACT

The unprecedented growth in the availability of data of all types and qualities and the emergence of the field of data science has provided an impetus to finally realizing the implementation of the full breadth of the Nolan and Temple Lang-proposed "integration of computing concepts into statistics curricula at all levels" in statistics and new data science programs and courses. Moreover, data science, implemented carefully, opens accessible pathways to STEM for students for whom neither mathematics nor computer science are natural affinities, and who would traditionally be excluded. We discuss a proposal for the stealth development of computational skills in students' first exposure to data science through careful, scaffolded exposure to computation and its power. The intent of this approach is to support students, regardless of interest and self-efficacy in coding, in becoming data-driven learners, who are capable of asking complex questions about the world around them, and then answering those questions through the use of data-driven inquiry. This discussion is presented in the context of the International Data Science in Schools Project (IDSSP) which recently published computer science and statistics consensus curriculum frameworks for a two-year secondary school data science program, designed to make data science accessible to all.

**Keywords:** International Data Science in Schools Project, accessibility, inclusivity, statistics education, data science education, data science in schools, computational thinking, statistics computation

## 1. Introduction

In 2010, Nolan and Temple Lang called for "computing concepts to be integrated into the statistics curricula at all levels." At the time of their paper data science had not yet emerged, yet their call for a "significant cultural shift to embrace computing" anticipated many current developments in statistics and data science education, including initiatives in statistics education that fully embrace computation (e.g., American Statistical Association Undergraduate Guidelines Workgroup, 2014) to full-scale collaborations between computer scientists and statisticians. The International Data Science in Schools Project (IDSSP) is one such collaboration, creating resources for data science education in secondary schools  (International Data Science in Schools Project

Curriculum Team, 2019). In this paper we describe the IDSSP and address and expand on the approach to computation adopted by the project.

IDSSP was initiated to address the needs of a rapidly-changing society in which leaders and citizens should have an understanding of how the use of appropriate methods for learning from data allow us to face important challenges, from the development of self-driving cars to addressing climate change, and also to motivate talented students to pursue further study in data science to address these challenges. IDSSP is a collaboration between statisticians, computer scientists, and educational experts in these fields.

The IDSSP team has developed curriculum frameworks (IDSSP, 2019) for both teaching data science in the final two years of secondary studies (assuming no prior knowledge in either computer science or statistics and no prior study of calculus) and for preparing teachers to teach data science. The focus of the frameworks is learning from data, and the development of the necessary computational skills and corresponding understanding are introduced as needed. A key IDSSP design criterion has been that data science learning in schools should be accessible to a very broad spectrum of students and not just the small minority for whom coding comes easily. It is critical that the community interested in the development of data science confront the danger that the paralyzing "math anxiety" problems that have bedeviled many approaches to statistics education might simply be replaced by "coding anxiety" in data science.

In Section 2, we describe IDSSP and how this desire for a broadly accessible curriculum influenced the consensus IDSSP approach to computation. Section 3 details this approach to computation, and Section 4 gives two case studies from topics in the IDSSP curriculum frameworks to illustrate how advanced topics in data science could be brought within reach of secondary school students when appropriate resourcing and scaffolding has taken place.

## 2. Context and Motivation

### 2.1 The International Data Science in Schools Project (IDSSP)

The International Data Science in Schools Project (IDSSP, http://www.idssp.org/) is an interdisciplinary project initiated and led by the ASA's 2019 Deming Lecturer, Nicholas Fisher, that involved an international team (with five countries represented on the Curriculum Team and additional countries represented on the advisory group) of both computer scientists and statisticians from the leading professional organizations for both disciplines. The project was motivated by a realization that the need for learning-from-data skills in our societies is vastly outstripping the ability of our slow-moving education systems to deliver, and by a desire to act to address that.

IDSSP wants data-science education to be valuable, accessible, enjoyable and enticing – conferring skills that are *valuable* for student's future lives, further study and careers; *accessible* to a broad spectrum of students (not just some small elite); *enjoyable* for both students and teachers; and *enticing* in the sense of arousing in students (and teachers) a desire to learn more.

The IDSSP Curriculum Team began by considering what would constitute ideal courses in data science at the junior and senior high-school levels. This was seen as a good place to begin because of its potential for broad impact. Having to build off a limited base of educator expertise in the school systems is unavoidable due to the rapid growth of data science as a field. As a result, it was recognized that the education of students cannot be addressed without also addressing the education of teachers, and the IDSSP curriculum frameworks include what we envision would be necessary to cover to train the teachers.

IDSSP was conceived as a project in two phases. Phase 1, completed in September 2019, was a scoping and goal-setting project culminating in consensus "curriculum frameworks" (IDSSP Curriculum Team, 2019), subsequently officially endorsed by many of the supporting societies, describing what is desirable to be included in a modern data science curriculum. (Note that "curriculum" is a dangerous word in an international context because it means different things in different countries). With help from core societies, project leaders have begun seeking financial support for a funded Phase 2 project with far reaching goals that include developing the resources to support courses based on the curriculum frameworks, and devising and implementing a course aimed at prospective teachers of data science.

The purpose of the IDSSP is to promote and support the teaching of introductory data science everywhere, not to be fine-tuned for any particular country. Because of the extraordinary variety of educational jurisdictions across (and even within) the various countries involved, and differences in patterns of prior educational exposure, it would be impossible to create a single course/curriculum that would satisfy all jurisdictional requirements. Accordingly, the curriculum frameworks developed, and the pedagogies and resources planned for Phase 2, are designed for flexible use by school systems and teachers for guidance and resources when preparing curricula and courses to meet their own local needs and challenges. Due to this, the IDSSP frameworks are much more detailed than might be expected for a high-level document. This is because very few people who will be reading and considering implementing them (in part or as a whole) will have a good overview of the broad sweep of data science and its various components. Consequently, the IDSSP Curriculum Team needed to lay out a fairly detailed map of the data science landscape to provide a useful picture of what is involved and what could be made accessible to students at the senior high-school level. It does, however, only provide only a map of a landscape to be traversed and not an ordered set of instructional sequences for traversing it.

In preparation for development processes in which statisticians and computer scientists can work together on a shared enterprise, IDSSP statisticians and computer scientists reached a common understanding of what a modern course in data science would most-desirably contain. This involved coming to understand one another's differing preconceptions and priorities and arriving at a consensus, an endeavour that inevitably involves compromise. As it turned out, we were pleasantly surprised at just how easy it was to arrive at shared perspectives. Apart from use of terminology, attitudes to computer programming were the biggest area of difference. To the computer scientists it was completely obvious that programming has to be a major component of "data science." In the Computer Science tradition programming, one of the major contributions of the area, is taught directly. Without question, programming skills and an appreciation of the power of programming had to be advanced even if actual expertise need not be a universal end-goal. For the statisticians, programming is just a means to statistical ends with a broad range of opinions about how much, if any, programming needs to be involved. Additionally, a fundamental tenet of IDSSP is accessibility for all students. Programming "requirements" could not, therefore be prescriptive. We had to make allowance for different strategies that could be applied to cohorts with different backgrounds. Articulating some of these enabling strategies is a major goal for this paper. We continue the discussion of the IDSSP approach to computation in Section 2.3.

## 2.2   Accessibility and Inclusivity in Data Science

The current demand for professional data scientists and the need for all citizens to understand the role of data in the decisions in their lives means that the educational system needs multiple pathways to the training of data scientists and practising data science. Data science problems require diverse perspectives and approaches to avoid bias and provide better solutions.  Key tenets of IDSSP are that our proposed introduction to learning from data should be accessible to all, and also inclusive.

A major IDSSP strategy for improving accessibility is to strive to minimise the extent to which failing to come to grips with one thing limits the ability to comprehend, or do, other things. This played out in a number of ways, such as emphasizing breadth of experience over detailed depth of coverage and in trying to minimise the extent to which modules build on other modules. It also manifested itself in the approach to computation (see Section 2.3).

We also want to eliminate barriers based on culture, experience or a sense of self-efficacy or belonging, that have historically kept talented students out of traditional STEM disciplines.  If our introduction is carefully designed, data science has the potential to be more diverse than other STEM disciplines as we can motivate learning with interesting problems from any area of interest to students.

Motivation is crucial to learning, influencing the "direction, intensity, persistence, and quality of learning behaviors in which students engage" (Ambrose et al., 2010, pp. 68-9).  Among the factors that have been shown to positively influence learners' motivation are problems they value as interesting and important (National Academies of Sciences, Engineering, and Medicine, 2018).  Moreover, harnessing this interest can make the need to learn technical content attractive to a diverse group of students. With the emergence of data science as a new discipline, we have the opportunity to harness its universal applicability to create a more inclusive STEM field, without the historical barriers to broad participation (Berman and Bourne, 2015; Lue, 2019).

In addition to working on problems they find interesting, students must also believe that they can and will be successful in order to be motivated to put in the effort to learn (Ambrose et al., p 76).  Thus we need to avoid barriers to self-efficacy that have kept some students out of traditional STEM disciplines.  Mathematics anxiety is a well-studied and widespread problem (Luttenberger et al., 2018), and concern has been raised about its negative role in producing sufficient graduates in STEM (Beilock and Maloney, 2015).  Similarly, statistics anxiety (Primi and Chiesi, 2018) and programming anxiety (Connolly, Murphy and Moore, 2009) have been identified as distinct but related constructs that may also contribute to negative student experiences and the avoidance of careers in data science.  An instructional strategy that has been shown to reduce math anxiety is increasing student motivation to learn by relating the material to students' interest and daily-lives (Luttenberger, Wimmer and Paechter, 2018), a strategy that is natural to learning data science.  Our proposed approach to computation in learning from data was designed with the goal of introducing computation as a valuable and powerful tool, and then building student confidence through teaching computational skills by stealth (see Section 2.4).

## 2.3 IDSSP's Core Computer Programming Goals

With the foregoing considerations in mind, a consensus decision was made that it was not the goal of the IDSSP frameworks to turn students who do not know how to program into competent programmers. Courses designed around the curriculum frameworks should be able to sit alongside programming courses in computer science with both possessing unique goals (as desired by the computer scientists).

This then begged the important question of what we wanted *all students* to experience with regard to computer programming in data science. The barest essentials, the team concluded, are:
- for students to gain an appreciation for the usefulness of code and its power for automating data science tasks;
- preventing fear of coding from ever taking hold;
- building confidence through starting to make relatively minor modifications to working code;
- and starting students on a journey towards learning to read code like a story, a sequence of human-understandable instructions, and towards learning to write their own code.

The IDSSP frameworks are largely agnostic about code use. Only the Topic Area in the frameworks entitled "The data-handling pipeline" focuses explicitly and systematically on programming ideas. But, for the rest, we envisage that the elements above can be introduced by stealth (see Section 2.4), so that experience and confidence gradually accrues. In subsequent sections we will discuss and illustrate some stealth strategies. The goal is: "This is not that hard. It is really useful and it can even be fun!" A key indicator of success would be a large proportion of students expressing a desire to gain some more skills in programming, and possibly pursuing further studies in computer science, mathematics or statistics thereby.

Data science, as IDSSP envisions it, is neither computer science nor statistics, and we deliberately aimed for the resulting curriculum to be neither, but something unique and novel. Furthermore, the decision was deliberately made (after extensive discussion between the computer science and statistics contributors) to focus on coding rather than computational thinking (in the quite broad computer-science sense of Wing, 2006, i.e., algorithmic thinking; problem breakdown into parts; non-coding algorithmic concepts), as having the most direct connection to our goal of learning how to learn from data (note that this does not preclude heavy GUI use in data science).

To further minimize unnecessary limitations on students' confidence to get started and access to capabilities, the IDSSP frameworks leave as many of the decisions about whether to use code (none/some/extensively) in most areas up to the choice of teachers, who are able to factor in the background and abilities of their students, as well as the availability of technology resources.

## 2.4 A Stealth Approach to Computation

Leveraging a focus on interesting problems and data-driven inquiry in data science, while powerful for motivation, is likely not sufficient to fully overcome anxiety and develop student confidence, a possible hindrance to accomplishing the above principles of our "barest essentials" for computation. Fundamentally, however, data science *requires* significant levels of computation. Whether by using dashboards, GUIs, or lines of code, one cannot work with data without somehow instructing a computer to perform some heavy lifting. It is simply not feasible to do practical, interesting, and engaging problems by hand in this field. Thus, to truly develop student skills, computation (but not necessarily coding) must have a central role.

As mentioned above, math and computer anxiety are powerful forces that must be combatted for student success. The IDSSP frameworks are designed around the ability to introduce computation and computational skills through *stealth*. By this we simply mean that we do not tackle something big and potentially scary (in particular, learning to write computer code) head on – at least not initially. Instead, initial learning happens tangentially in the process of trying to do other things. The ignition of motivation and interest in students can drive their desire for skills to improve their ability to better tackle the interesting problem(s), and then the required skills gained organically via student-driven desire and need.

In general, good teaching is often a stealth activity (Sharp, 2012). Students' curiosity and desire to learn can be encouraged by providing motivation through interesting problems that align with their interests or for which teachers have ignited an interest. To solve such problems, students then need new skills.  Learning objectives related to the acquisition of these skills can be disguised through activities such as games or problems from another area. This stealth motivation has been used in a variety of settings to teach computational and quantitative reasoning. Yevseyeva and Towhidnejad (2012) motivated the development of computational thinking through the analysis of a recent nuclear accident in a secondary school chemistry class; Shreve (2005) describe the use of computer games to motivate learning; and Gunn (2017) describes the teaching of quantitative methods in political science through the study of election data.

Data science is a natural environment for stealth learning of computer programming skills as we have bigger purposes in play (understanding data) and we don't necessarily have to write code to be able to do the majority of data understanding (e.g., quite powerful visualizations can be done entirely via GUI, and projects like iNZight (Wild, 2019, www.inzight.nz) demonstrate that even modeling can successfully be done in this way as well - see 3.2.3 for examples). In this way, learning concepts-and-skills for data analysis and learning to work with code can occur in parallel with very little constraint on the relative speeds at which each of these threads has to be advanced. This leaves these settings open to teacher or system choices, taking into account the background and inclinations of their cohort of students.

As a very simple example of stealth, consider a student trying to make a graph. The teacher provides a function call that the class can run (in an environment that makes changing and running code trivial) and students are led to observe that it produces something close to what they want. The function call has arguments that are obviously tweakable, and tweakable in obvious ways. So they can play with changing things, rerunning the code and observing what happens. Some changes will be useful for the purposes of the desired graph, while others might just be fun (e.g., inappropriately vibrant colours from a data visualization perspective, yet still appealing to a new learner). Students will largely not be aware that they are on a learning journey with computer code. They are just playing with pictures. Over time, students' feelings that "this is not doing quite what I want it to do" can provide motivation to take a little class/lecture time to learn something that will overcome the issues. Working with code recedes from being a feared "other" to just an accustomed way of interacting with computers. There is no substantive research that we are aware of about implementing such stealth strategies well but a good deal of anecdotal evidence from colleagues that this can work.

The overarching goal of the project is to teach people to learn from data motivated by uncovering the stories data can tell. Computation plays an indispensable role in pursuing this overarching goal. As learners encounter data that is sufficiently large or complex, the need for and convenience of computational tools becomes readily apparent. Students can begin with tools that facilitate gaining insights from the data quickly, but over which they have minimal control. As their sophistication increases and they can experience the value of code to produce analyses and, when they are motivated to customize their output, they can progress to working more directly with code and gaining broader and deeper experience with what they can accomplish through code-based approaches. In every respect, we propose that learning be motivated by problems that are interesting to students (a challenge to the perceptiveness of teachers) and also make the need for computational tools self-evident, helping students to realize that computational skills give them *power*. Increasing complexity of problems leads to the need for more sophisticated tools and abilities, and these can be developed in a gradual, scaffolded fashion: GUIs → code that they can modify and rerun (system generated or teacher-provided, e.g., in Jupyter (Kluyver et al., 2016) or R Markdown (Allaire et al., 2020) notebooks) → writing more substantial pieces of code (which could be high/intermediate/low level, as discussed further in Section 3.2.3).

## 3. Building Computational Skills

The IDSSP frameworks were designed to provide access to many valuable areas that have never been part of the mainstream of introductory statistics or computer science. There is much more emphasis on data acquisition, privacy and ethics, data provenance, data wrangling techniques and the whole data-handling pipeline. Coverage of visualization techniques (including interactive and dynamic graphics) is much more extensive and multivariate than in normal introductory treatments. Exposure to many new areas is provided for by including selections from introductions to supervised and unsupervised machine learning, geolocated (map) data, seasonal time series data, text mining and analytics, image data, and recommender systems. It would simply be impossible to experience much of this in a reasonable timeframe if students were required to write detailed code!

To learn data science, we believe students must *do* data science, and as discussed above, this requires a variety of levels of computational skill, depending on the desired outcome. In this section, we outline our motivations further, and describe some specific examples to provide clarity as to possible perceived implementation paths. Recall that the project chose to be explicitly code-language agnostic, with computer scientists and statisticians alike agreeing that specifying languages or even classes of languages was too restrictive for a curriculum framework. The thought was that any jurisdiction that chose to use the framework and develop courses on top of it would make decisions based on their local resources and availability of training for educators.

### 3.1 Accessibility and goals for computation

Modern software can contribute enormously to making data science concepts accessible to a broad range of students and for equipping the students with substantial capabilities to apply these concepts in practice to interesting, real, complex data. The project's view is that using computers running modern software is fundamental to any serious exposure to data science. We recognize that this raises serious issues of access to technology (a problem COVID-19 lockdowns have highlighted), but such issues have been postponed to Phase 2 (described in Section 2.1).

However, access to modern computers and software is not the only issue of accessibility we confronted in this project. Even with a highly-resourced school environment, superbly trained instructors, and students deeply invested in a desire to learn from interesting data, any approach to implementing a successful course that engages a broad set of students must be cognizant of the need for all students to feel that they are capable of success.

As the IDSSP goal that trumps all others is facilitating the ability to learn from data, it is possible to leverage the very wide variety of capabilities that can be accessed through GUI systems with no need for code. When code is used, we advocate heavy use of high-level functions. The typical objective for programming in computer science courses is programming mastery, so that the student can write code from scratch, given a task description – typically beginning with small, easily understood tasks. In such a framework, the eventual result is long sets of instructions solving small problems. By comparison, most programming done by data science practitioners makes heavy use of calls to high-level functions written by others, so that succinct and easily translated (code to natural language or vice-versa) commands result in a great deal of sophisticated processing. Data science education needs to work with high-level instructions (as practitioners do) because we need our students to become capable of obtaining serious insights and accomplishing powerful results early, so as to support their motivation and willingness to believe in their own learning capabilities. This is much more difficult to accomplish using lower-level programming approaches. Data science students also need to know that "do it yourself" is a very bad practice when well-tested software exists. It is not just an inefficient use of time, but also leads to large numbers of unnecessary errors.

In Section 3.2 we expand on our vision for how coding skills can be developed in this context.

## 3.2 Exemplar Pathways for Coding Growth

One emphasis of the IDSSP curriculum framework is the encouragement of repeated tasks or situations where students are provided a skeleton or template demonstrating a technique, and encouraged to "do something different". Rather than being a teacher-directed task ("create a boxplot"), students will hopefully be encouraged to explore and learn and be the architects of their data-driven analyses. This integrates nicely with the Data Science Cycle, where posed problems lead to data, and upon extensive work, eventually loop back to adaptation of problems and further data gathering.

The use of GUI structures (Section 3.2.1) and system-generated code (Section 3.2.2) in parallel naturally allow for evolution in skill on the part of students, starting at the point-and-click level of data analysis, and through careful, judicious use of system-generated code, slowly removing training wheels and working into modification of pre-written and provided code strings. This, if taken to its natural conclusion, should eventually result in students who have gained sufficient mastery of the concepts to be able to write their own snippets of data analysis code (Section 3.2.3), fulfilling the core programming goals discussed in Section 2.3. Furthermore, this progression has no set time scale attached, so one implementation for a topic area could simply never leave the GUI framework, whereas another (perhaps with a prerequisite of a computer science class) might leap immediately into system-generated code and aim higher.

The use of notebooks (Jupyter or RStudio (RStudio Team, 2020)) and R Markdown files for reproducible analyses is also a natural medium for introducing code re-use and mastery, by providing learners with an already functional literate document (Knuth, 1984). Reading and understanding what such a document does can then be scaffolded to allow for modification ("change a variable") or extension ("what if we consider this other factor?"), and is a beautiful way to drive motivation toward coding through stealth.

Depending on the emphasis of courses designed in this philosophical vein, an additional area which connects is that of pipelines, especially in the informatics or data analysis sense. Sufficiently motivated students may be guided toward the use of more complex data sources which are decidedly *not* tidy, which will in turn necessitate the introduction of sophisticated multi-stage data wrangling techniques. Taken to its professional conclusion, this is a data processing pipeline, and students could slowly be motivated to understand the necessity of such pipelines by connecting the "blocks" of munging and analysis together, and demonstrating that doing such work as a monolithic blob just leads to chaos.

## 3.2.1 GUIs as a Vehicle for Growth

Into this discussion, we would like to add some cognitive realities about human memory. Short term memory is a severely limited resource; studies reported by Cowan (2000, 2001) suggest that the average person can only hold two to six pieces of information in their attention at once. Long-term memories for the details of how to do things fade fast when they are seldom used. This is a particular problem for programming. But both programming and the use of many GUI systems run into the related problem of knowing and remembering names. You cannot do anything until you know, and remember, the name of the thing you want to do. This is a significant barrier to getting started and also results in significant time-loss getting back up to speed after a period of inactivity.

Some of the biggest advantages of GUI systems flow from the visual cues and reminders they provide of "What's on offer here?" Interfaces can provide a structured, top-down way of encountering a new area (e.g., working with network data) with progressive revelation guiding thinking through a problem via context-aware choice sets. Thus GUIs can provide serious new capabilities quickly with reduced learning curves, with their memory cues reducing the dependence on human memory. Thus, they are well placed for presenting options for "How else can I *look at* that?" (display types), and "How else can I *do* that?" (methods) with instant delivery of results from each tentatively-entertained option – leveraging the modern tendency of people, when given buttons and choices, to ask "I wonder what that does?" and just try things out. This is not to say that many existing GUI systems use these capabilities well. But good GUI systems are helpful for beginners, occasional users, and doing one-off things really fast (provided the system prioritizes them). They can enable users to see a whole range of things we can do with our data and do them very quickly and with very little effort. Furthermore, the clicking, dragging and hovering gestures inherent in a well-designed GUI can also create a level of immersion in data which is quite different from the (often cold) remoteness of coding systems. This is particularly powerful in the hands of relative novices[1].

The biggest deficiency of GUis is their inability to do new things the system has not allowed for. The *flexibility* and *extensibility* that coding provides is probably the biggest reason we need to start students on their journeys towards coding. There are many others. We have already mentioned the *automation* of repetitive tasks. The ability to *reuse* and *share* code, as-is or after modification, confers huge time-efficiency and knowledge-transfer benefits. This is really brought home by the "Oops! Do it again!" experience: an involved analysis has to be repeated because a mistake was discovered in the data. Sharing code facilitates *reproducibility* – the ability for someone else to reproduce (and therefore check) an analyst's results. It automatically generates an *audit trail* of what was done and how, and allows *disciplined practices* for working, sharing, version control, and creating *dynamic documents* in which expository text is interspersed with blocks of code that generate the desired data-analysis outputs such as tables and graphs within a report and can easily be recompiled if changes are needed. Students can profitably experience quite a bit of this, so the benefits of working using code are brought home for them, but without any need to become experts.

## 3.2.2 System-generated code

GUIs and coding have complementary strengths, and students may benefit from exposure to both. But there are also GUI systems that can write code, making available the code that implemented the actions that the point-and-click interface requested. This code can be taken away to be used elsewhere or modified and rerun in the system. Perhaps the best known and most long-standing example of this in the R ecosystem (R Core Team, 2020) is R Commander (Fox, 2016) which is a graphic user interface for R. Blue Sky Statistics (www.blueskystatistics.com), Jamovi (www.jamovi.org/) and RKWard (https://rkward.kde.org/) also have this ability; see Muenchen (2020) for comparisons.

---

[1] We thank one of the reviewers for this excellent point.

The systems above appear to be aimed primarily at data-analytic practitioners and their interests are not pedagogical. IntRo (Hare and Kaplan, 2019) is an R Shiny app produced for intro-stat courses that makes available the R code it uses to generate its results (Hare and Kaplan, 2017). It is limited to the analyses most often used in standard introductory statistics, but is also much simpler for beginners for that very reason.

Large parts of iNZight (Wild, 2019), and also its online version iNZight Lite, construct and make available R code. The system takes user instructions from the GUI, constructs R code to implement them, and then runs that code and also stores it (automatically for anything that changes the data, and otherwise by user request). Very recently, in the interests of further facilitating strategies being discussed here, iNZight's technical lead Tom Elliot has designed and implemented the model exemplified in Figure 1.

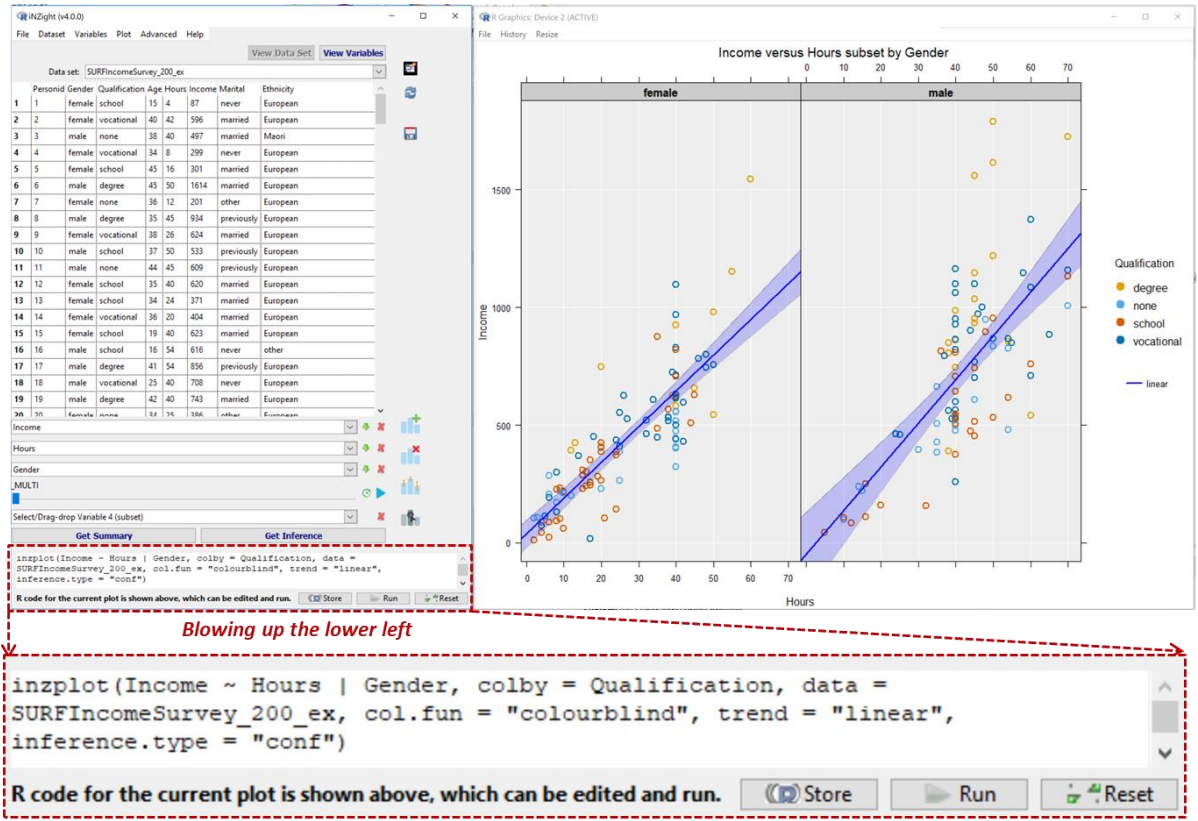

**Figure 1:** *Relationships between GUI settings, code and output in iNZight.*

In all of the windows most often used by beginners:
- The function call that produces the display you have just asked for is shown – with provision for storing, or changing and rerunning the code
- Settings in the GUI determine the function call and the output
- Changing and rerunning a valid function call changes the settings in the GUI to match your code (**Reset** returns things to the last set of GUI instructions)

The strategies in play are:
- "the code that does it" is always in view to foster learning by osmosis
- The mappings between GUI settings, argument values of the function call and output are direct, to foster seeing the relationships between them
- Because the system is responsive, each request in the GUI for a minor change or customisation triggers an instant change both in the output and in the code, also helping highlight what particular code elements do
- Opportunity is provided to experiment with the code within a familiar environment with expected behaviours and expected outputs
- Restricting what is always being shown to just the current function call keeps things simple and makes the mappings between code and GUI settings more obvious. (Other strategies are required for learning to "string together" code.)

Currently code is stored automatically for all data-wrangling operations performed, and when asked, for graphics and statistical output produced by the base module (including ggplot code) and the generalised-linear-modeling module.

System- (or GUI-)generated code provides an alternative to instructor-provided code as a source of building blocks for a program. Take data-wrangling operations as an example: we would not expect students to become *expert* at data-wrangling. Even if it was possible, the time it would take would be completely out of proportion to its value in the time available for an introduction to data science. But we would at least like students to often experience using data wrangling operations necessitated by particular data sets they were working with as part of repeatedly experiencing the whole data-science cycle; a version of the cycle (IDSSP Curriculum Team, 2019, pp 8-9) is an anchoring scaffold for the frameworks. Note that Nolan and Temple Lang (2010) also reinforce the importance of working through the whole investigative process.

iNZight has the capacity to use dialogs to lead users through almost all of the data-wrangling operations in the book *R for Data Science* (Wickham and Grolemund, 2016) and write the R code needed. This means that when students need to employ particular operations to wrangle their data in preparation for analysis, the system and its documentation can lead them through those operations, even with side-by-side pre- and post-views of the data for the more complicated ones, and then provide the code they need for their own program to automate a larger process. Working this way contributes to understanding operations and using them in a program. And it can be done at the individual (rather than class) level. Students can also process and analyse data using a GUI, obtain the code for a program that duplicates everything they have just done, and then modify that program to adapt it for a somewhat different purpose. Obviously, this can also be done with pre-written reproducible documents (e.g., R Markdown or Notebooks; Jupyter notebooks), which students can take and run in order to see functionality, and then later modify for new data sets or new explorations.

We have seen no substantive pedagogical research on using system-written, or teacher-provided code to help with learning to code, just data analysts talking online about how the code-writing provisions of R Commander (for example) helped them to learn R. There is a need for research not only into how code generated by existing systems can be used to enhance learning but also on the ways code delivery and interaction in statistical and data-science GUI systems can be structured to better help learners.

## 3.2.3 Levels of coding

One of the beauties of the modern interpreted code ecosystems (e.g., R (R Core Team, 2020) or Python (Python Core Team, 2020)) are that they allow exposure to coding at many levels of programming, from students writing their own direct-data-modifying instructions to calling incredibly complicated frameworks like TensorFlow for machine learning. In the following, we discuss parallel implementations of the same objective at different levels, to demonstrate this capability, notebook using the R language. Similar presentations could be done using Python, e.g., beginning at a high level with pre-implemented frameworks and Jupyter notebooks; progressing to specifying functions more directly in the intermediate stage; and concluding at the low-level stage with base Python and, perhaps, a library such as matplotlib.

In the following we describe an example problem from three different perspectives: very high-level code such as might be provided from a GUI-driven system; intermediate code such as might be written in a data science framework such as the tidyverse (Wickham, 2019) or mosaic (Pruim et al., 2017); and low-level code such as might be written by a more experienced programmer who is *still* a data science novice. Having programming experience does not translate to data analysis immediately, and we expect that at least some of the implementations of the curriculum will be designed for students who have programming experience but little-to-no data experience, and no statistics background.

We now elaborate on our distinction between higher level versus lower level coding, echoing the distinction between high-level and low-level programming languages. Coding using *tidyverse* commands ([www.tidyverse.org](www.tidyverse.org)), for example, tends to be higher-level than programming using *base R* commands in the following sense. The tidyverse contains collections of powerful functions, built on top of base R commands. The tidyverse commands are intended to make some very important and commonly used operations much easier for an analyst to perform, with less code and less attention to detail than is needed with base R commands.

This higher-versus-lower-levels-of-coding distinction is important for strategizing about tackling the problem of knowing and remembering names introduced at the beginning of Section 3.2.1. With code you cannot do anything unless you know and remember the names of the things you are trying to do, and the syntaxes used in putting them together. Lower levels of coding require more programming complexity and higher demands on human knowledge and memory, thus reducing broad accessibility to students.

Our brief illustrations start with the same small, clean rectangular data set from a workforce survey read into a dataframe called **incomeData**. The first 6 rows are:

```
    Personid Gender Qualification Age Hours Income Marital Ethnicity
1          1 female        school  15     4     87   never European
2          2 female    vocational  40    42    596 married European
3          3   male          none  38    40    497 married    Maori
4          4 female    vocational  34     8    299   never European
5          5 female        school  45    16    301 married European
6          6   male        degree  45    50   1614 married European
```

***Very high-level code*** *(in the sense that almost everything about how things are done is decided by defaults)* This example uses functions from the iNZightPlots package.

    inzplot(~Income, data=incomeData)

"Show me the data on the variable *Income* (using a plot)."

    inzplot(Income~Gender, data=incomeData)

"Show me the relationship between *Income* and *Gender*."

    inzsummary(Income~Gender, data=incomeData)
    inzinference(Income~Gender, data=incomeData)

"Tell me more about this relationship using (i) the summary statistics and (ii) the inferential information that people usually want to see in a situation like this."

How these commands are "obeyed" is automated using defaults in the software (taking variable type into account). The simplicity of code that does not expose details makes it very easy to see how important things can be changed; in this case it is easy to see how other variables can be explored and what you would need to do to use a different data set. This also provides a good starting point for learning to read code as a narrative. R is well suited to this as function arguments with defaults need not appear in function calls.

In the above there is no control of what happens and how. We can get more control by overriding defaults and/or by moving to lower levels of coding. We pay a price, of course, in complexity.  By adding arguments that override defaults, the calls above (particularly the plotting calls) are also enormously customisable, thus opening up a huge range of opportunities for exploring the data using features such as changing plot types, faceting, sizing and colouring elements by other variables and labelling; all with little programming complexity, e.g.,

```
inzplot(Income~Qualification | Gender, data=incomeData, plottype="gg_violin")
```

With extremely high-level code like this, where small pieces of code deliver rich information, students' first experiences of direct interactions with code can associate it with power, not drudgery. This is very similar to the philosophy espoused by Çetinkaya-Rundel (2018) of the importance of the first-day hook: get students up and running with powerful, interactive data analysis as early as functionally possible, so that the power overwhelms the fear.

***Intermediate code*** *(in the sense that some decisions about how things are done are decided by defaults, but the user is expected to provide a little more guidance)*

We lower the coding level slightly with this example by using the mosaic package from Project MOSAIC (Pruim, Kaplan and Horton, 2017) to obtain similar output.

```
gf_boxplot(Income ~ Gender, data = incomeData)
favstats(Income ~ Gender, data = incomeData)
confint(t.test(Income ~ Gender, data = incomeData))
```

We now need to know a little bit more to issue our requests. We need to know that the type of plot we want to see is called a boxplot, the name of the function that gives you that particular plot, and that for our inferential information we want confidence intervals from the t.test function (this last instruction is not actually mosaic).

Mosaic is an important package for learning journeys in R. From its inception, it has prioritised enabling consistent syntax, and easily readable code that conveys a narrative, as a means of stripping away unhelpful complications from students' coding journeys.

***Lower-level code*** *(in the sense that fewer decisions about how things are done are decided by defaults, and the user is expected to almost fully guide the analysis)*

Our final example lowers the code level further with a call to ggplot2 and some base R code to produce summaries and inferential information that, for this particular combination of variable types, are almost the same as produced by our highest level of code above.

```
ggplot(data = incomeData, aes(x = Gender, y = Income)) +
      geom_boxplot() +
      geom_dotplot(binaxis = "y", dotsize = .5) + coord_flip()
male_incomes <- incomeData$Income[incomeData$Gender == "male"]
```

```
        female_incomes <- incomeData$Income[incomeData$Gender == "Female"]
        c(summary(male_incomes), sd(male_incomes, na.rm = TRUE))
        c(summary(female_incomes), sd(female_incomes, na.rm = TRUE))
        t.test(male_incomes)$conf.int
        t.test(female_incomes)$conf.int
        t.test(Income ~ Gender, data = incomeData)
```

The core point we are trying to make here is that by using sufficiently high level functions we can reduce the complexity of the code students have to work with immensely, and consequently the cognitive and memory demands put on them. This gives the opportunity to build coding experience gradually without sacrificing speed of progress through learning-from-data experiences even in an entirely code-driven course. Topic areas encountered early in the course sequence can be tackled with simple calls to powerful high-level commands, while topics encountered later may be able to make more detailed and complex coding demands as students accumulate experience and confidence in working with code. Both ends of the spectrum are illustrated in the Case Studies in Section 4.

In computer programming terms, everything we have shown above is, actually, still very high-level programming. Hadley Wickham's ggplot is a very high level function with powerful abstractions which facilitate building complex multi-part graphs in a principled way. Even R is itself a very high-level programming language. What we have really been illustrating here are just shadings of degree.

## 4. Case Studies on Learning Computation by Stealth from the IDSSP Modules

In Section 3, we discussed how GUI-driven code can be scaffolded to provide novice students with a starting point which will be accessible to almost any computer user. We then examined and defined different "levels" of coding expertise, categorized primarily based on the amount of support and lifting that the computer system provides. Now, to illustrate the IDSSP approach to computation, we provide two case studies in how teachers and students might engage with topics in Unit 2 of the IDSSP curriculum frameworks: working with time series data, and interactive data visualization. Each case starts with consideration of a rich, multivariate data set, albeit of different formats, to motivate the need for acquiring new skills and to inspire students' desire to creatively explore. For these topics, students will need sophisticated computational ideas that are not traditionally taught at the secondary school level. The case studies illustrate how our approach to teaching computational skills by stealth can make these topics accessible to all students. In addition, following from Section 3.2.3, the two case studies use quite different coding levels, as a demonstration of the fact that the depth of the material and exploration does not need to be closely tied to the coding level *if* the computational tools available allow for high- or intermediate-level coding.

Our first Case Study works with seasonal time series data. It takes a high-level approach easily accessible early in a coding learning-journey. It uses very simple, high-level function calls to the iNZightTS package. Because these are also the function calls that iNZight's GUI-system itself issues and there are obvious mappings between the code and settings in the GUI, the GUI and code could even be used in parallel if so desired. This case study was actually written to be a literate document using R Markdown, and has been brought into this paper in its present form for demonstration purposes - we envision that a classroom setting might start with a templated markdown framework showing a worked example, and then successively adapt that framework into something like what we have here.

The second Case Study on interactive data visualization, coming from the computer science perspective, takes a more intermediate- or low-level approach in the sense that the coding experience is much more central to, and integrated with, the statistical experience. Similar to the first case study, the data visualization work was originally created as a Jupyter notebook, and a similar classroom vision applies. Please note we are not arguing here that time series should be approached one way and interactive visualisation the other. Both can be approached from anywhere on the spectrum between GUI and low-level code, and we are simply illustrating how two very different choices can play out.

## 4.1 Case Study: Climate over time in Toronto

The data-science educational context for this case study is learning to investigate and learn from seasonal time-series data. The real-world context is global warming (or "climate change"), a topic of great interest to many young people. Wherever you live on our planet, the **climate** of our environment appears to be changing. NASA's Earth Observatory program says "The world is getting warmer. Whether the cause is human activity or natural variability - and the preponderance of evidence says it's humans - thermometer readings all around the world have risen steadily since the beginning of the Industrial Revolution … the average global temperature on Earth has increased by about 0.8° Celsius (1.4° Fahrenheit) since 1880. Two-thirds of the warming has occurred since 1975, at a rate of roughly 0.15-0.20° C per decade."

> ***Case Study:*** *Given that we consider "local" to mean Ontario, Canada (as four of the authors of this paper live there!), (1) Can we observe a warming **trend** in* *the data from Toronto, Ontario, Canada?; and (2) What **prediction** can we make about average temperatures in the near future?*

We will explore some temperature data from Environment and Climate Change Canada which are **time series**: repeated observations of the same unit or measurement over **time**. ECCC maintains hourly observations of temperature at many locations across Canada which are freely available on a data portal at http://climate.weather.gc.ca/. Similar databases exist for other countries such as Australia, New Zealand, the United States of America and the United Kingdom as well as most other countries on Earth.

A primary motivator for developing a computer program for this situation (and almost all time series data) is the fact that new data is always accumulating as time marches on, so we are very likely to want to update any analysis we do sometime in the future to include all the new data. Our process is to first obtain the data from the database, then wrangle it to get it into a form that the R package we want to use likes as input, and then start analysing it.

The database allows us to extract hourly, daily, or monthly data. We will take monthly data. We get a rectangular data set with the variables/fields shown in Figure 2 if we request data for Toronto's Pearson International Airport:

> Longitude (x), Latitude (y), Station Name, Climate ID, Date/Time, Year, Month, Mean Max Temp (°C), Mean Max Temp Flag, Mean Min Temp (°C), Mean Min Temp Flag, Mean Temp (°C), Mean Temp Flag, Extr Max Temp (°C), Extr Max Temp Flag, Extr Min Temp (°C), Extr Min Temp Flag, Total Rain (mm), Total Rain Flag, Total Snow (cm), Total Snow Flag, Total Precip (mm), Total Precip Flag, Snow Grnd Last Day (cm), Snow Grnd Last Day Flag, Dir of Max Gust (10's deg), Dir of Max Gust Flag, Spd of Max Gust (km/h), Spd of Max Gust Flag

**Figure 2:** All fields available from ECCC for climate data from Pearson International Airport, in the format presented when downloaded from the portal mentioned.

There are fields that are redundant for this analysis because they are characteristics of Pearson Airport and never change, e.g., latitude and longitude. There are others that we do not need such as "Mean Temp Flag" which takes the value "M" if the value of "Mean Temperature" is missing, and is blank otherwise; a little investigation reveals there is no missing data from 1970 (actually after 1937). And we may also narrow our focus to a smaller number of variables of particular interest. So we want a wrangling step to reduce the number of variables being considered to a smaller number of variables of particular interest.

When the data is imported, R does not like special characters in variable names; e.g. "Extr Min Temp (°C)" is automatically converted to "Extr Min Temp..C." So another desirable wrangling step is to change the variable names after import to simpler, more attractively-readable names. Additionally, the R package iNZightTS we will use for our time series analysis in this case study does not automatically recognise year-month data in the form "1971-02", but it does automatically recognise "1971M02" so we want to change the format used by the Date/Time variable (which turns to "Date.Time" on import).

Once we have wrangled our data we have a data set ready for initial analysis. But when we come to update the analysis with new data we would not want to do all this again. If we were to have a program that does all the steps, and we want to update the analysis we would only have to re-run the program. Coming up with the code to do these things will be either beyond most students, or take a long time to arrive at, and is well beyond the scope of an introductory course. Additionally, different data sets require different sets of wrangling steps (from minimal to excessive). Adapting the opening sentence of Tolstoy's *Anna Karenina* (about families), all tidy data is alike but every untidy data set is untidy in its own way. So in this context, teacher input to code formation will normally be required. Alternatively system-written code could be helpful.

Once we are in the tidy world, with variables organized, everything is much better systematized. (Please note that, in what follows we emphasize the connection between commands and what they produce, regardless of the programming environment.)

**head(pearsonClim)**

```
      date year meanTemp meanMaxTemp meanMinTemp totalPrecip maxWind
1 1970M01  1970    -10.9        -5.9       -15.8        25.7      61
2 1970M02  1970     -6.6        -1.6       -11.6        29.7      72
3 1970M03  1970     -2.3         1.9        -6.4        43.7      85
4 1970M04  1970      6.6        12.3         0.9        81.3      90
5 1970M05  1970     12.0        18.1         5.9        56.6      63
6 1970M06  1970     17.1        23.9        10.3        38.1      60
```

**library(iNZightTS)**
**meanTemps = iNZightTS(pearsonClim, var = "meanTemp")**
**plot(meanTemps, t=100, xlab = "Time", ylab = "Mean Temperature (C)")**

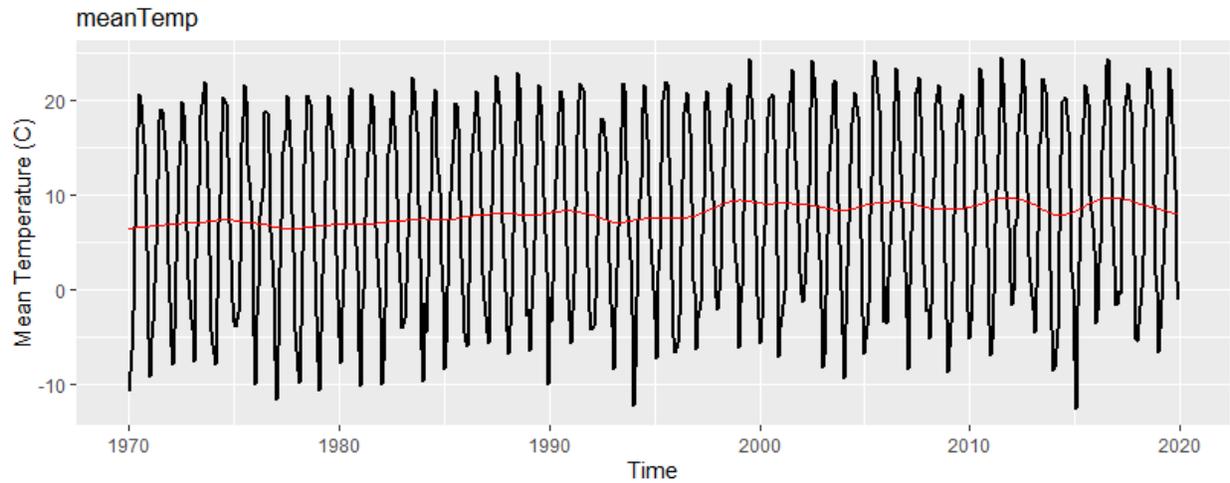

```
seasonplot(meanTemps)
```

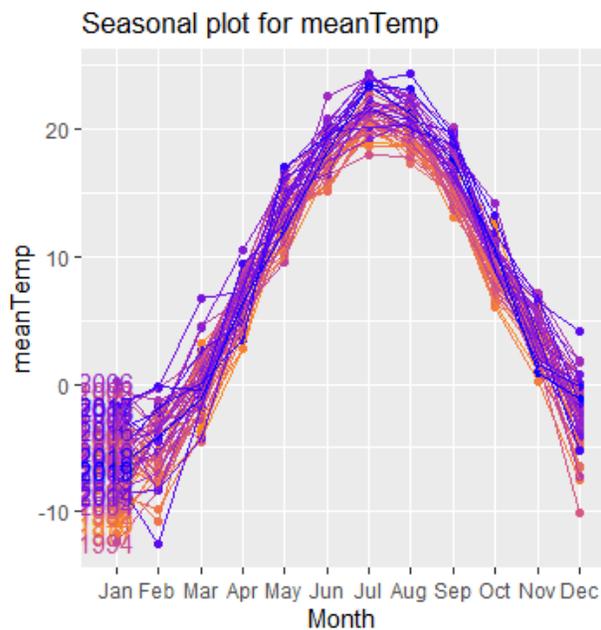
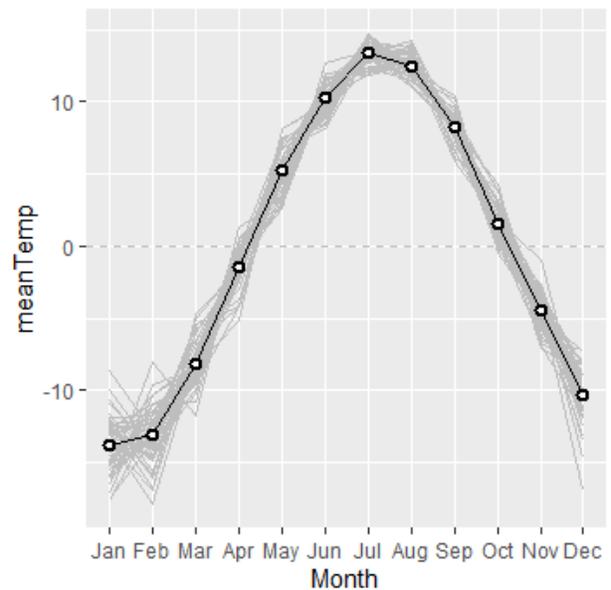

**Figure 3:** Display of the available variables after cleaning and tidying the data, and an introductory graphical analysis of the mean temperature series, from 1970 to 2019.

```
plot(meanTemps, forecast = 36)
```

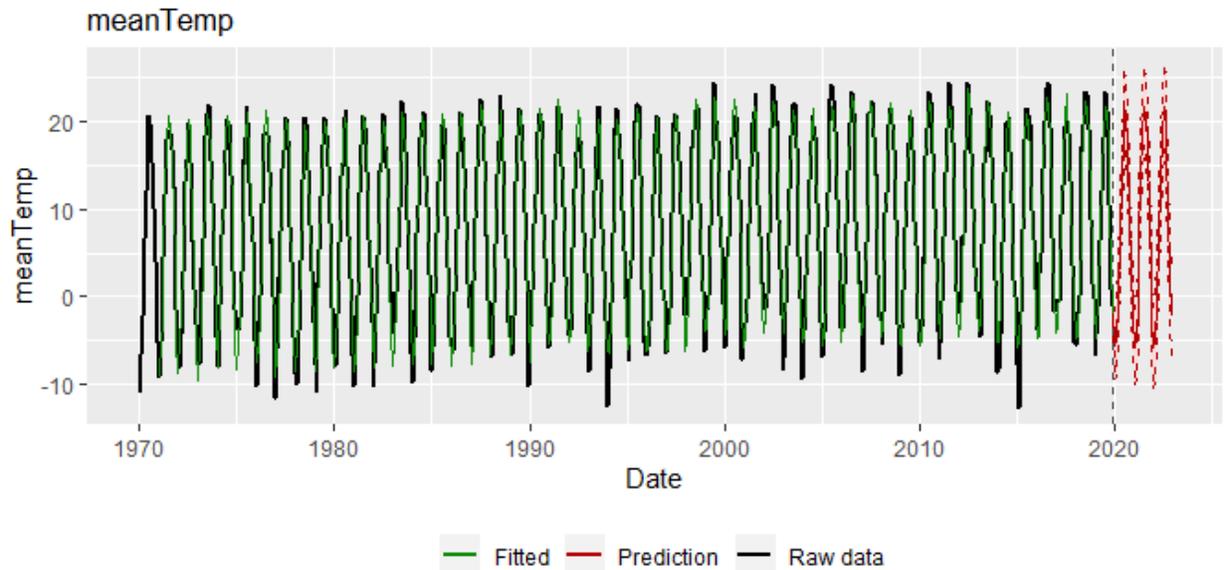

Forecasts with upper and lower confidence limits

```
              fit        upr        lwr
Jan 2020  -5.6143922  -1.6224643  -9.606320
Feb 2020  -4.8569001  -0.8380746  -8.875726
Mar 2020  -0.3247944   3.7207499  -4.370339
Apr 2020   6.3629026  10.4349903   2.290815
... ...       ...         ...        ...
Oct 2021   9.7061322  14.2778520   5.134412
Nov 2021   3.0950695   7.6902944  -1.500155
Dec 2021  -2.0812654   2.5373450  -6.699876
```

**Figure 4:** Forecasting mean monthly temperatures for the next 3 years (36 months)

Figure 4 shows that the seasonal movements (month of the year effects) in the mean temperatures are much larger than any movement in the trend (as we would have expected). Accommodating the seasonal differences restricts the depiction of movement in the overall trend to a very small interval of vertical space on the graph, thus making it much harder to see what is happening to the trend. So next we aggregate our data to yearly averages and plot that.

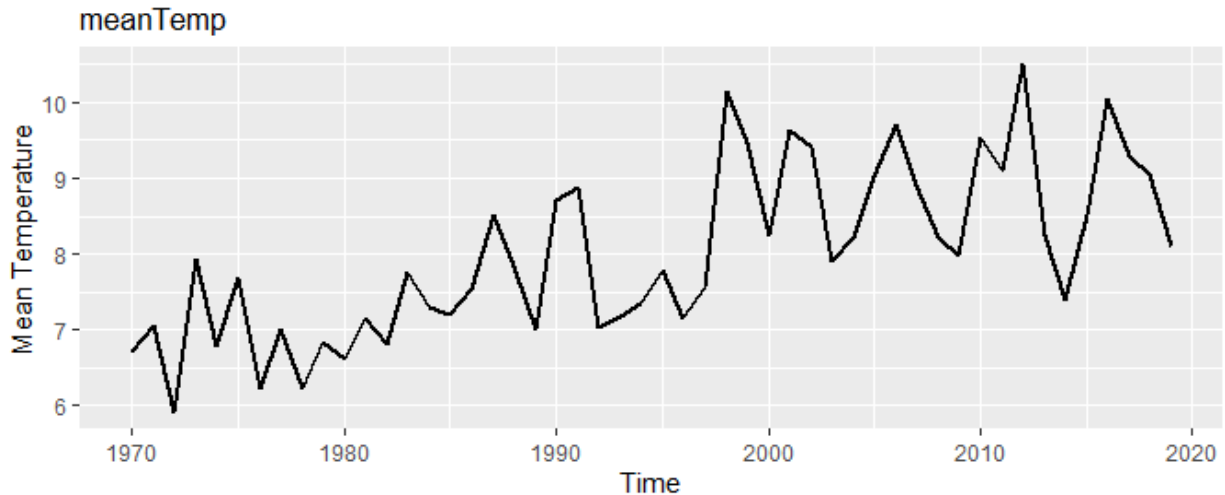

*Figure 5*: *Mean yearly temperatures at Toronto Pearson International Airport, 1970-2019.*

In Figure 5 a warming effect is now much more obvious. Mean annual temperatures at Toronto Airport appear to have climbed fairly steadily from 1970 to around the year 2000 but seem to have leveled off since.

Other variables in the data show contrasting behaviour, for example, total precipitation and maximum windspeed have very weak seasonal effects and do not show any consistent pattern of increase or decrease over the time period.

The types of plot shown above, together with decomposition plots and plots of multiple series, have been the mainstay of teaching, data exploration and assessment for time series at the last-year of high school in New Zealand since 2013. (In NZ, the majority of students entering their last year of school study statistics.) The educational experience is centred on using graphs of data to make sense of the world and the students typically use GUI systems (Census at School NZ, 2020). High-stakes assessments call for written reports on what the student has learned from a particular set of data.

Although we have just shown a minimalist computational-approach to time series here, time series data is a natural fit as a vector for computation by stealth since so much of available real-world data is actually recorded as time series. Thus, going back to our earlier point about motivation: if teachers wish to encourage student-driven learning via motivated investigations, they will almost certainly encounter time series data as the natural fit for their question-driven inquiry. Additionally, time series come with fascinating structural behaviour that can be visible directly from plots, allowing students to feel powerful as they "detective" their way through an investigation. Further, time series naturally come with a date-time structure, which almost *requires* code to be easily wrangled: encouraging the cleaning of dates and times is a great hook to get students interested in repeatable, reproducible analysis-focused code.

## 4.2 Case Study: Interactive data visualization

Many modern data sets are highly multivariate, meaning items in the data set have many attributes, and these attributes can be quantitative or categorical. Many such data sets are represented in a tabular format in a standard relational database. Examples include the features of a computer (speed, storage, price, screen resolution), the characteristics of a movie (length, box office sales, date of release, genre), or information about athletes' performance (name, sex, age, height, weight, year, sport, medal). While tables are a helpful way to organize and sort data, discovering relationships between attributes using a table or a collection of tables can be challenging. Appropriate visualizations of such data can help identify patterns or form hypotheses (Anscombe, 1973; Fekete *et al.*, 2008). However, because of the sheer number of attributes, it is not trivial to determine which particular visual representations of high-dimensional data sets best support decision making.

This case study gives an approach to computation for interactive data visualization by stealth through a learning sequence. It assumes the data set is complete and does not have significant errors, as dealing with data cleaning and missing data is a different (typically prerequisite) problem. There is a plethora of open multivariate data sets that are readily available online, and teachers can also engage students in producing a data set of their own through simple data collection methods involving sensing technology, surveys, or even directly scraping data from webpages where allowed, or through APIs. This allows for endless possibilities in terms of the themes and problems that students can perform data analysis on, to foster motivation. Most importantly, teachers should focus on a data problem that is relatable to the students, for which the answer is not trivial, and for which the data set is rich and complex enough to pose interesting data analysis and visualization challenges.

> ***Case study:***
> *Your parents are considering purchasing a new computer for the family. You have done some research and found a spreadsheet of about 6,000 models to consider, including the price, processor speed, hard drive size, RAM, graphics capabilities, brand reputation, and warranty. Can you trim down this list to retain 5 top options based on their properties? How can you then effectively communicate to your parents why these options are best compared to any other computers in the initial list?*

The data set used in this use case is large both in terms of the number of items and dimensions to consider, making it virtually impossible to process the data manually. Note that teachers could also easily adapt this case to identifying a subset of points of interest in any multivariate data set and situating these items with regards to the larger collection (e.g., What are the best cities to travel to? What book should I read next? What are the most endangered species in North America? What are the highest ranked colleges? What makes an exceptional athlete exceptional?).

The first activity with the computer data set in this case study is to create single charts using commercial GUI-driven visualization technology, such as Tableau, PowerBI, or even Excel to explore different facets of the overarching question students are trying to answer. For example, the students could examine the relationship between *price* and *processor speed* using a 2D scatterplot. Given the complex relations between the attributes of the data set, and the number of items, it will quickly become apparent that examining the data from one point of view at a time does not give a clear picture of the options. This activity is targeted at strengthening accessibility through positive self-perception, i.e., it is not only the ability to write code that matters. By building up confidence gradually through engaging with data using well-tested software, the students will gain proficiency in plotting but also will gain data literacy through discovering new questions that cannot be answered with a single static chart. Through this activity, students should also learn the lesson that concurrently considering the multiple dimensions of a multivariate data set often requires not only one, but multiple visual representations, each of which provides a different perspective on the same data. Teachers can encourage students to think critically about the particular visualization or collection of visualizations that could be generated to support decision making, and discuss the difficulty for the analyst or viewer to integrate and combine the information from these multiple static views.

The next activity will be to create **coordinated views** of the data, sometimes called a **dashboard**. These views can be interactively linked to reveal the same data item across plots, or to highlight items matching a selection on any given plot. For example, a scatterplot of *price* and *processor speed* can be interactively linked to a histogram of the number of computers in each bin of *RAM* (see the first plot in Figure 6 for the histogram) to allow for interactive highlighting of points in each *RAM* group. This activity builds the ability of students to reason about data from multiple perspectives, seeing that a data item (a computer in this case) can be viewed in different ways based on the question of interest. Commercial software such as Tableau and PowerBI offer the capability to create these dashboards in a lab activity. Because there are virtually an infinite number of ways to design a single dashboard, learners will again face the question of what particular set of views should be integrated to their own dashboard.

This is a good spot for initiating a discussion about *analytical goals* and *tasks*, and how visualization supports these. For instance, teachers can ask students to walk the class through the reasoning process that led to selecting the best five candidates using their dashboard. Having multiple students expose their analysis would allow them to compare and contrast the set of questions and results that the different dashboards created by each (group of) student(s) support. Teachers should help learners understand that while there are better choices in terms of dashboard design, there is no definite answer as to what is the best dashboard. A dashboard that excels at supporting certain analytical tasks will inevitably be less good at supporting some other tasks. An important lesson learned through this activity is that visualization design involves making compromises, and that different solutions can be equally valid, depending on where analytical priorities are set.

From a computational thinking perspective, the above activities using commercial software will build the capacity to see data and visualizations of data as separable and modifiable, with the software creating **representations** of the data.

Dashboards and coordinate views are powerful tools to perform **exploratory data analysis**, that is, dynamically querying the data by making selections and filtering data, then observing patterns in the different views. Exploratory data analysis in an interactive tool like Tableau or PowerBI allows students to investigate a variety of questions, make discoveries, and form hypotheses. However, they are not the best medium to communicate insights to stakeholders, as they require a viewer to identify important take-aways by interacting with an interactive tool without further guidance.

The **communication of insights** requires the analyst to curate one or a collection of visual representations of the data that best illustrate the data-based evidence that supports the message to be conveyed to enable decision making, often also accompanied with explanations in writing that make it clear to the intended audience what the important information is.

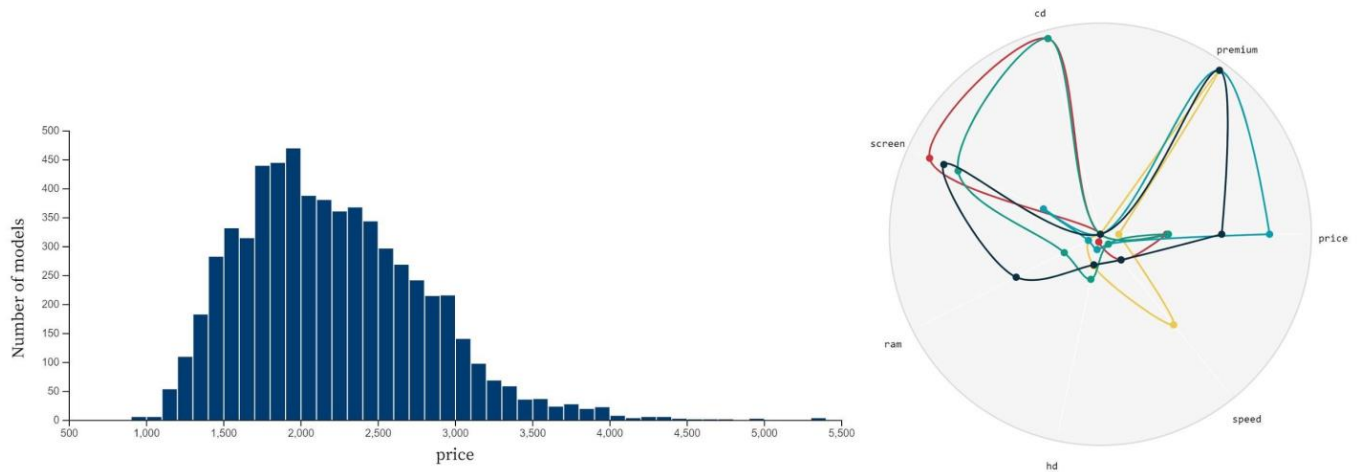

**Figure 6:** Histogram displaying the number of computer models binned, and a radar chart of the features of the 5 selected computers, developed using Observable notebooks.

Having narrowed down the options to a few favorites using Tableau, the student will be asked to create a "report" for their parents comparing the characteristics of their favorite 5 options, so that the pros and cons of each can be communicated. A radar chart is one way to compare multiple attributes for a limited number of items. Implementations of the radar chart are available online to **run** and **modify**. First, the student will examine an existing radar chart (e.g., https://observablehq.com/@palewire/radar-chart and compare with the right-hand plot in Figure 6) using a notebook-style environment called Observable, which is built atop d3 and is easily accessed online. At this point, the student may be getting their first exposure to programming through a literate programming environment (Knuth, 1984), which allows the exploration and running of existing code snippets with little-to-no setup. Other options include R+Shiny or Python+Jupyter Notebooks. Stepping through code written by others, or making low-risk minor modifications and observing the outputs will allow the students to experiment without the pressure of "writing code".

The next activity will be to **change the data set** from the one used in the example to the extracted data set from a previous related exploratory activity. In the development of computational thinking, this helps students understand the separation between code, its visual outputs, and the underlying data set. A second activity will be to run a different code snippet to **encode** the data in a different way, for example, using a parallel coordinates plot (e.g., https://observablehq.com/@miralemd/parallel-coordinates-plot, and Figure 7). The goal of this step of the activity is to illustrate that the same data set can be represented in different ways. A radar chart creates a 'shape' for each computer which can be compared, while a parallel coordinates plot more clearly illustrates which item is at the top or bottom on each feature.

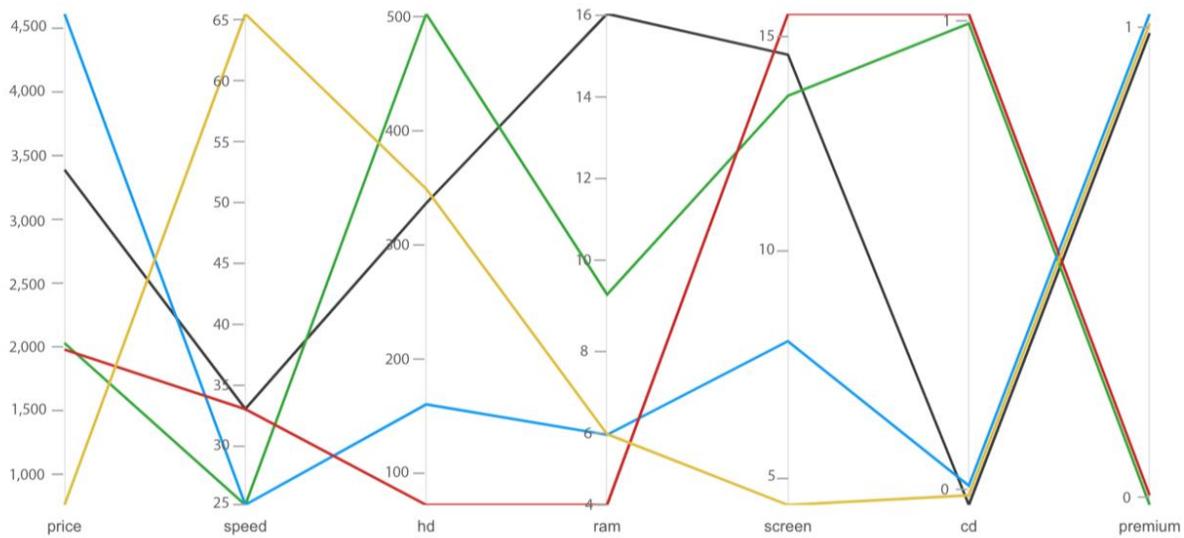

**Figure 7:** Parallel coordinates plot showing the characteristics of all 5 selected computers.

The student will now be asked to **modify presentation** aspects of the code to create a better report to give to their parents. Through this activity of styling a visualization, the students will be exposed to color models, rendering styles, and scaling functions. Examples of changes to try include changing the color scheme, the line style, the scaling of each axis, or the fill of the curves in the radar chart - no expectation of mastery, but exposure often inspires students who, while not comfortable coding, still are familiar with the breadth of visualizations from their environments. In addition, labels, and legends should be added as appropriate.  This activity builds on the ability of the student to interpret existing code as a sequence of instructions to process and create a visualization from the data given as input to these functions and develop their ability to refer to the documentation of functions to identify what each function does, and what are each of its parameters. Our goal is for the students to understand that, besides the actual data that is given as an input, a function often encompasses other input parameters that allow them to customize a more general concept, as for instance, "plot". The student may then appreciate that the function has some default values for most of these optional parameters, but can be customized to their needs, such as making axes visible or not in a chart, or changing the title. Further, the students may learn that there exists a plethora of pre-existing, well-tested code that is well documented, that they can reuse for their own purpose.

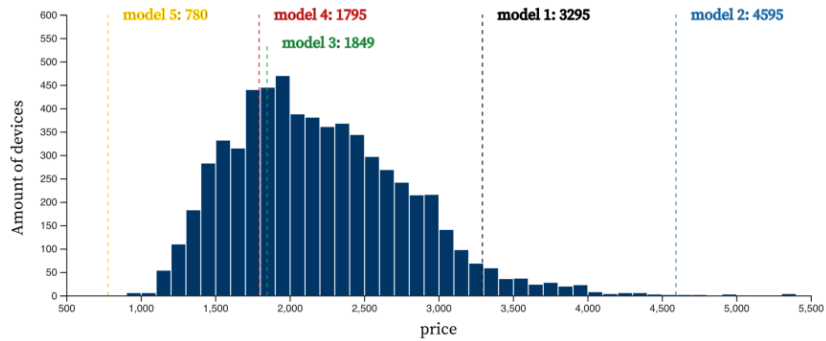

**Figure 8:** An improved histogram displaying the number of computer models binned, customized to contextualize the price of the five selected models compared to the whole collection of computer models.

Now the scenario is extended to imagine that parents respond that they are concerned the student has selected expensive computers based on brand rather than making a balanced choice across capabilities and price. Countering this argument could be achieved by situating the selected computers within the big picture of the broader data set. Thus, the student is tasked with adding a distribution of attribute values across the complete data set using a set of histograms, one for each attribute. While answering the scenario-driven question, the student is introduced to **repurposing code** by splicing together code snippets from other types of charts, such as histograms into the same workbook. From a computational thinking point of view, this activity introduces the interoperability of code snippets and the modularity of software, through putting parts together to achieve a deeper analysis. This will expose the student to **built-in functions** in programming languages that are already wrapped-up to perform common operations, such as aggregating data across a dimension into a distribution.

The student aims to show the price is reasonable while achieving high values on the target dimensions. For this, the student would need to **calculate derived measures** from the values. The new view should annotate the mode, median, and percentiles of each attribute. Once the code is created for one attribute, students will learn that it can be **replicated** across the other blocks of code. The workbook will now allow a reader to contrast the actual values of the suggested best computers against the global distributions of attribute values. Here, the student is building proficiency on statistical measures and how to calculate them, while also being exposed to computational concepts such as iterative loops and array data structures as a byproduct. Again, the intention would not be for such students to master loops - that material more properly belongs in a distinct computer science course - but to be exposed to them, and to gain comfort with the underlying ideas.

At this point, there is still a disconnect between the selected computers illustrated in the radar chart and the distributions depicted in the histogram. To bring these together, the student will be tasked with **adding standard interactions** to help the parents understand the data. Items selected from the radar chart should be highlighted in the summary histograms, i.e., show the bin the item falls in across the distribution (Figure 8). In reverse, bars in the distributions could be selected to filter ranges on the radar chart. Further, axes in the parallel coordinate chart should be re-orderable, and brushing values along one axis should result in filtering data whose value for that dimension falls within the brushed range. Here, through the goal of creating visuals for clear communication, the students are learning the computational concept of real-time modifiability through interaction.

In order to use their workbook to create a convincing purchasing rationale, the module should be completed by having students add explanatory text blocks to help their parents understand the analysis which was carried out, and how to use it.

## 5. Conclusion

In 2010 as one of the dimensions to ensuring continued and broader impact of statistical sciences, Nolan and Temple Lang called for statistics educators to expand the use of computation in statistics curricula at all levels. They identified three key components to the integration of computing more broadly in statistics: (1) Broaden statistical computing; (2) Deepen computational reasoning and literacy; and (3) Compute with data in the practice of statistics. Ten years on, the emergence of data science, and the corresponding integration of computing and statistics in the pursuit of learning from data has brought focus and urgency to this call, and gave impetus for data science, including these components, to be brought to more students and to start earlier than post-secondary school. The IDSSP curriculum frameworks have all been designed with (1) and (2) as implicit goals: statistical computing is necessary and required for any reasonable approach to learning from data, and reasoning from data and computational literacy are dominant members of the curriculum, in an approach that allows learning data science to be accessible and inviting to almost all secondary school students.

A traditional statistics course may find a course built on IDSSP to be somewhat foreign; similarly, a computer science course could also feel some culture shock. This is by design: data science is not exactly statistics nor computer science, and lives in an uneasy balance between the two. The three key components of Nolan and Temple Lang presaged these changes that are now required to translate a traditional pen and paper statistics curriculum into something suitable for the modern era of learning from data. All aspects of the IDSSP frameworks are designed with the motivation of gaining skill in learning from data, in the structure of the cycle of data science, including formulating a problem, acquiring, exploring and analyzing the data, and communicating the results.

Almost without exception, everything in the IDSSP frameworks is based on heavy use of software tools in the pursuit of extracting learnings from data (Nolan and Temple Lang's component 3). Obvious exceptions are thinking and learning about: ethical considerations, what data can/are and can't/aren't telling us, and verbal communication. Anything even moderately close to the breadth of coverage provisioned for in the frameworks would be impossible without excellent, and occasionally bespoke, high-level computational tools. For most areas existing software should suffice; for others new software may need to be written to ensure sufficient accessibility. But we believe this is all perfectly doable using the software tools we already have (e.g., in the R and Python ecosystems) and adapting software models already in existence (including notebooks and IDEs). IDSSP was initiated to try to speed up progress in the spreading of data-science thinking and skills; to move faster than business-as-usual educational processes have been capable of delivering. This requires a little entrepreneurial spirit and risk taking, and will require a lot more hard thinking and hard work from many people, perhaps including readers of this paper, to help bring aspirations to fruition.

The design of the IDSSP curriculum frameworks, especially in the explicit inclusion of computation from almost the very start, proposes a path for how we might begin the development of these components in secondary school for a broad range of students. The hope is that we will encourage the pursuit of advanced study in data science for a group of students who may not have been previously motivated to do so and increase understanding of data-based decision-making for all students. Engagement is key in order to gain this retention, and we believe that learning coding via stealth, and developing an understanding of the power that comes with it, is a key weapon in teachers' arsenals as they attempt to ensure just that. Interesting questions in a variety of areas can be included to engage the students and motivate them to learn technical material when needed, with need-driven learning leading to students obtaining technical and computational skills "just in time", as their motivation and interest leads them to understand and communicate the information that can be learned from the data.

The advent of interactive visualization tools and literate programming frameworks has the potential to change the restriction of "writing code" being a gatekeeper, and allow students with varying experiences and diverse interests to engage in and be able to *do* basic data science. This is fundamentally a question of accessibility, and our hope is that teaching computational skills through stealth will allow greater access to students from a broad range of backgrounds and interests. Furthermore, investigating data from a variety of contexts can allow otherwise disengaged students to be motivated to learn, and, we hope, to see a pathway for themselves in data science, a question of inclusivity. Encouraging agency in learners has tremendously compounded benefits in engagement and retention, and ensuring breadth in the academic skill, background and interest of data science students will encourage the growth of the data-driven questioning framework in disparate fields, and eventually, if perpetuated, across all of society. Long-term, this inclusivity may also help our professions tackle fundamental questions like algorithmic bias which are presenting themselves in our modern, data-driven economy.

Unfortunately, there is currently limited research available on the explicit benefits and possible detriment of what we have proposed, including the stealth approaches and more particular considerations, for example, how the development of computational thinking in this context can be supported starting with GUIs. We would like to draw the reader's attention to the need for research in these areas, as the consensus opinion of the IDSSP curriculum team was that these methods have great promise for the development of accessible learning in data-science computation in comparison with others we considered.

While accessible learning with data and computation is at the core of the IDSSP frameworks, beyond the consensus that computation be treated as a scaffolded, introduction via stealth, activity, the Curriculum Team intentionally decided to refrain from stating "a right way to do computation." Quite different approaches can be used to achieve very similar ends, and suit different teachers and different types of students. We have touched on some possible computational approaches in this paper. We have discussed why and how the IDSSP frameworks leave choices around computational tools to teacher, or education-system agency. But whatever computational approaches are chosen, IDSSP's desired ends cannot be met unless the groups of students being taught find their computational experiences unthreatening, interesting, empowering and enjoyable.

The emergence of the field of data science has transformed the landscape of data-driven inquiry across our world. The IDSSP framework of learning from data, in the structure of the cycle of data science, has been designed to build on this change, with an admittedly utopian goal that data science should be accessible to all, regardless of either mathematical ability and interest, or of computational experience and confidence. The world has become incredibly complex, and we need citizens of our societies to be able to ask correspondingly complex questions, and then be able to answer them!

Pruim, R., Kaplan, D. and Horton, N. (2017), "The mosaic Package: Helping Students to Think with Data Using R," *R Journal* 9, 77-102. DOI: 10.32614/RJ-2017-024

Python Core Team (2020), "Python: A dynamic, open source programming language," Python Software Foundation, available at https://www.python.org/.

R Core Team (2020), "R: A language and environment for statistical computing," R Foundation for Statistical Computing, Vienna, Austria, available at https://www.R-project.org/.

RStudio Team (2020), "RStudio: Integrated Development for R," RStudio, PBC, Boston, MA, available at http://www.rstudio.com/.

Sharp, L.A. (2012), "Stealth Learning: Unexpected Learning Opportunities Through Games," *Journal of Instructional Research*, 1, 42-48.

Shreve, J. (2005), "Let the Games Begin. Video Games, Once Confiscated in Class, Are Now a Key Teaching Tool. If They're Done Right," George Lucas Educational Foundation, available at https://www.edutopia.org/video-games-classroom.

Wickham, H. (2019), "tidyverse: Easily Install and Load the 'Tidyverse'," R package version 1.3.0, available at https://CRAN.R-project.org/package=tidyverse.

Wickham, H. and Grolemund, G. (2016), *R for Data Science: import, tidy, transform, visualize, and model data*. O'Reilly Media, Inc. Available at https://r4ds.had.co.nz/.

Wild, C. (2019), "*iNZight (VIT, Plots)*: A simple data analysis system which encourages exploring what data is saying," available at https://www.stat.auckland.ac.nz/~wild/iNZight/ and https://github.com/iNZightVIT/.

Wing, J.M. (2006), "Computational Thinking," *Communications of the ACM*, 49, 33-35.

Yevseyeva, K. and Towhidnejad, M. (2012), "Work in Progress: Teaching Computational Thinking in Middle and High School," in *Proceedings of the Frontiers in Education Conference*. DOI: 10.1109/FIE.2012.6462487
38